\documentclass{iau_FM}
\usepackage{graphicx}

\title[Standardization in the UV with Astrosat and its issues related to star cluster studies] 
{Standardization in the UV with Astrosat and its issues related to star cluster studies}

\author[Priya Shah]
{Priya Shah}

\affiliation{Department of Physics, Maulana Azad National Urdu University, 
Gachibowli,\\
 Hyderabad 500 032, India \\ email: {\tt priya.hasan@gmail.com} \
 }
 
\pubyear{2018}
\jname{Calibration and  Standardization Issues in UV-VIS-IR Astronomy
} 
\editors{Susana Deustua ed.}
\begin{document}

\maketitle

\begin{abstract}
The Ultra-Violet Imaging Telescope (UVIT) is one of the payloads in Astrosat, the first Indian Space Observatory. The UVIT instrument has two 375 mm telescopes: one for the far-ultraviolet (FUV) channel (1300–1800 ˚A), and the other for the near-ultraviolet (NUV) channel (2000–3000 ˚A) and the visible (VIS) channel (3200–5500 ˚A). We shall discuss the issues with standardization in the UV with reference to Astrosat Observations (Cycle A04). I shall discuss the problems faced in data-analysis and how these in turn lead to serious issues dealing with the color-magnitude diagarms, membership and age of the young embedded clusters studied. 

\end{abstract}

\firstsection 
\section{Introduction}

Astrosat is India’s first dedicated multi wavelength space observatory and was launched on Sep 28, 2015. One of the unique features of Astrosat is that it enables the simultaneous multi-wavelength observations of varied astronomical objects with a single satellite. The payloads (telescopes) observe in the visible, ultraviolet and x-ray region of the electro-magnetic spectrum. They are the The Ultraviolet Imaging Telescope (UVIT), Large Area X-ray Proportional Counter (LAXPC), Soft X-ray Telescope (SXT), Cadmium Zinc Telluride Imager (CZTI) and Scanning Sky Monitor(SSM).

The UVIT consists of two 375 mm telescopes a far-ultraviolet (FUV) channel (1300-1800$\AA $),
  near-ultraviolet (NUV) channel (2000-3000$\AA$) and the visible (VIS) channel (3200-5500  $\AA$).  It  provides simultaneous imaging in the two ultraviolet channels with spatial resolution better than $1.8''$, along with a provision for slit-less spectroscopy in the NUV and FUV channels (http://uvit.iiap.res.in/). 

\section{Target clusters}  
  \cite[Camargo et al. (2015, 2016)]{cam2} used WISE data to identify clusters in high latitudes of the galaxy and  2MASS data to find the cluster parameters after careful decontaminations procedures were followed (Table \ref{sobs}) \footnote{$A_V$in the cluster central region,  age, from 2MASS photometry, $R_{GC}$ calculated using $R_{\odot}$ = 8.3 kpc as the distance of the Sun to the Galactic centre,$x_{GC}$, $y_{GC}$, $z_{GC}$: Galactocentric components}.

\begin{table}[h]
\caption{Fundamental parameters and Galactocentric components for the ECs (\cite[Camargo et al. 2015]{cam})}

\label{sobs}
\small
\begin{center} 
\begin{tabular}{| l |l| l | l |l|l|l|l|}
\hline
Cluster &  $A_V$ &   Age   &  $d_{\odot}$ & $R_{GC}$ & $x_{GC}$  & $y_{GC}$  & $z_{GC}$   \\
        & (mag)  &   (Myr) &  (kpc)       &  (kpc)   &  (kpc)    &  (kpc)    &  (kpc)\\ \hline
C 438  &  0.99 $\pm$ 0.03& 2$\pm$ 1& 5.09 $\pm$ 0.70& 8.69 $\pm$ 0.40& -07.04 $\pm$ 0.02 & +0.97 $\pm$ 0.13& -4.99 $\pm$ 0.69 \\
C 439  &0.99$\pm$ 0.03 &2 $\pm$ 1 &5.09 $\pm$ 0.47& 8.70 $\pm$ 0.26 & -07.05 $\pm$ 0.02& +1.06 $\pm$ 0.10& -4.97 $\pm$ 0.46\\       
C 932   &1.40$\pm$ 0.03& 2 $\pm$ 1&5.7 $\pm$ 0.53&10.55 $\pm$ 0.29&-9.07 $\pm$ 0.17&-0.29 $\pm$ 0.03&-5.38 $\pm$ 0.50\\
C 934&1.46 $\pm$ 0.06&2 $\pm$ 1&5.31 $\pm$ 0.51&10.27 $\pm$ 0.27&-8.97 $\pm$ 0.17&-0.27 $\pm$ 0.03&-5.01 $\pm$ 0.48\\
C 939&1.30 $\pm$ 0.06&3 $\pm$ 2&5.40 $\pm$ 0.50&10.34 $\pm$ 0.27&-9.00 $\pm$ 0.17&-0.31 $\pm$ 0.03&-5.09 $\pm$ 0.47\\
C 1074&0.93 $\pm$ 0.06&3 $\pm$ 1&4.14 $\pm$ 0.39&9.12 $\pm$ 0.15&-8.18 $\pm$ 0.09&-2.66 $\pm$ 0.25&3.02 $\pm$ 0.28 \\
C 1099&0.71 $\pm$ 0.06&5 $\pm$ 1&4.32 $\pm$ 0.61&7.32 $\pm$ 0.30&-6.03 $\pm$ 0.17& -3.61 $\pm$ 0.51&2.05 $\pm$ 0.28 \\
C 1100&0.93 $\pm$ 0.06&1 $\pm$ 1&6.87 $\pm$ 0.36&8.00 $\pm$ 0.23&-4.76 $\pm$ 0.13&-5.59 $\pm$ 0.29&3.16 $\pm$ 0.16 \\
C 1101
&0.96 $\pm$ 0.06&3 $\pm$ 1&3.91 $\pm$ 0.55 &6.83 $\pm$ 0.27&-5.78 $\pm$ 0.20&-3.16 $\pm$ 0.44 &1.78 $\pm$ 0.25 \\    

\hline

\end{tabular}
\end{center}
\end{table}

\begin{figure}[h]
\includegraphics[width=10cm,height=5cm]{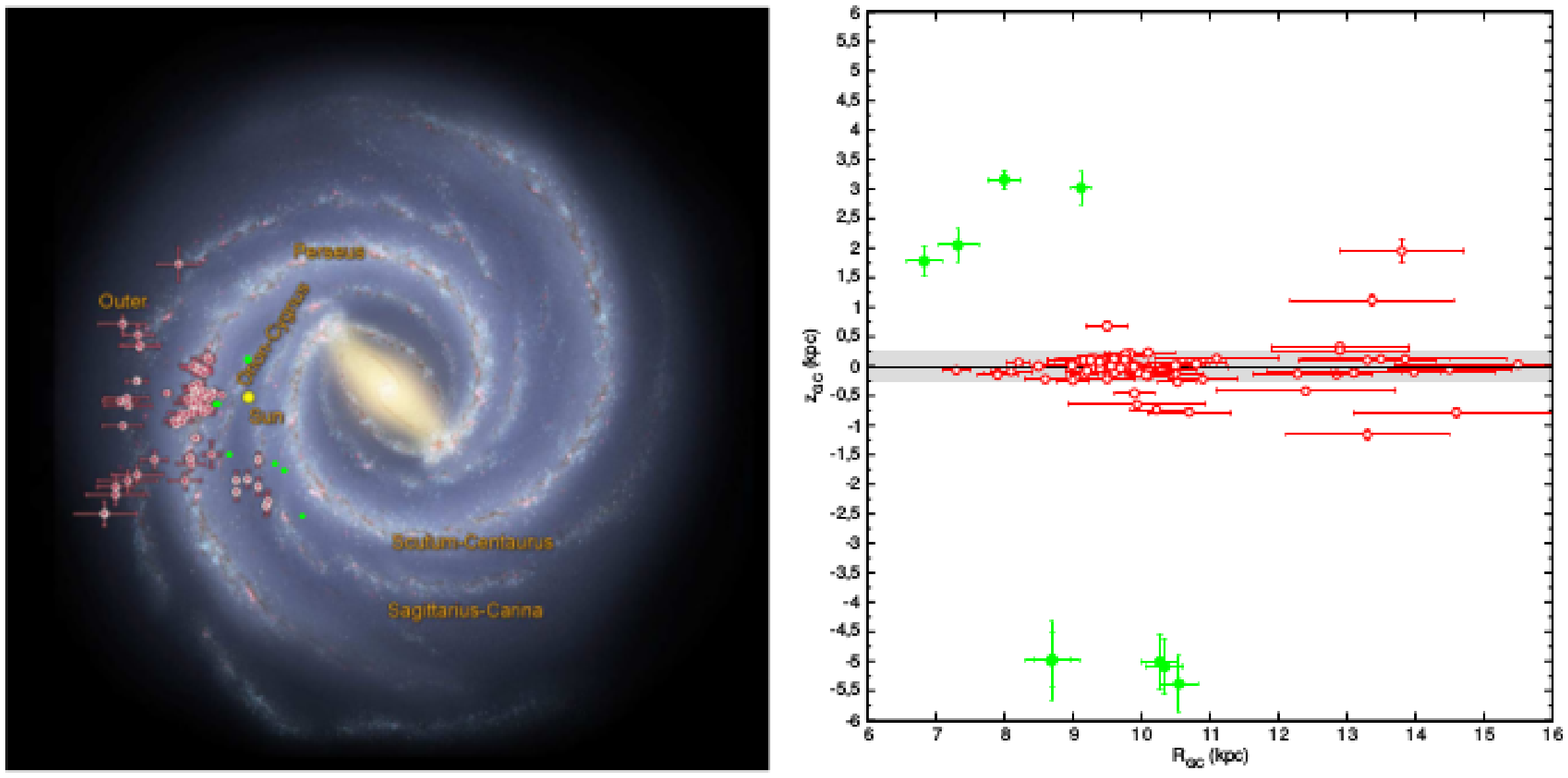}
\caption{Spatial distribution of the ECs in this study (green circles) compared to ECs in  previous works (red circles). Credit: Robert Hurt (NASA/JPL) and Camargo et al.. (2015).}
\label{clus}
\end{figure}

Figure \ref{clus} shows the spatial distribution of the newly found clusters (green circles) compared to the earlier studies (red circles) by Carmago et al. (2015, 2016). 

The discovery of these high latitude clusters are very crucial to our understanding of the galactic halo. If these young stars are formed in the halo, then it is possible that these clusters may get unbound before they reach the disc and young stars may reach the disc isolated. We also need to assess if this is an episodic event or a regular feature. 

There are two possible scenarios that can explain star formation at such high galactic latitudes. One possibile scenario could be Galactic fountains or infall. 
The expansion of substructures powered by massive stellar winds and supernovae can trigger star formation in various shells and rings, inputting energy to the superbubble (\cite[Lee et al. 2009]{lee}). 

The other possible scenario is extragalactic in nature. The Milky Way galaxy has several satellite galaxies in its vicinity. Tidal interactions of the galaxy with its satellites is also a possible reason for star formation to take place so far from the disc of the galaxy. There are 12 known satellites of our galaxy.

\section{Observations}
We proposed simultaneous observations  of these clusters using UVIT and the Xray telescopes on Astrosat, the Indian Astronomy Satellite. We shall concentrate only on the UVIT data. Our proposal A04-080 was granted  a total observation time of 4500 secs where we observed TPhe (calibration source) and the two clusters C438, C439. We observed in the FUV (Filter: 2 - Barium Fluoride for 300 secs, Filter: 3 - Sapphire for 1200 secs) and NUV (Filter: 3 - NUV13 for 300 sec, Filter: 2 - NUV15 for 1200 secs). 

 \cite[Postma et al.(2011)]{2011post} describe calibration data and discuss performance of the photon-counting flight detectors for the UVIT.  \cite[Tandon et al.(2017)]{2017tan} reported on the  performance of the (UVIT) on-board AstroSat. cite[Murthy et al.(2017)]{2017mur}) wrote a  software package (JUDE) to convert the Level 1 data from UVIT into scientifically useful photon lists and images. The routines are written in the GNU Data Language (GDL) and are compatible with the IDL software package. The level 1 data was analysed using the UVIT pipeline as well as the the JUDE pipeline, we found reasonable agreement with the two but were unable to construct good color-magnitude diagrams for the clusters. 
 
\section{Results}
We used Gaia DR2 \cite[Gaia Collaboration et al. 2016]{2016A&A...595A...1G} data to study our clusters.
\begin{figure}[h]
\includegraphics[width=15cm,height=5cm]{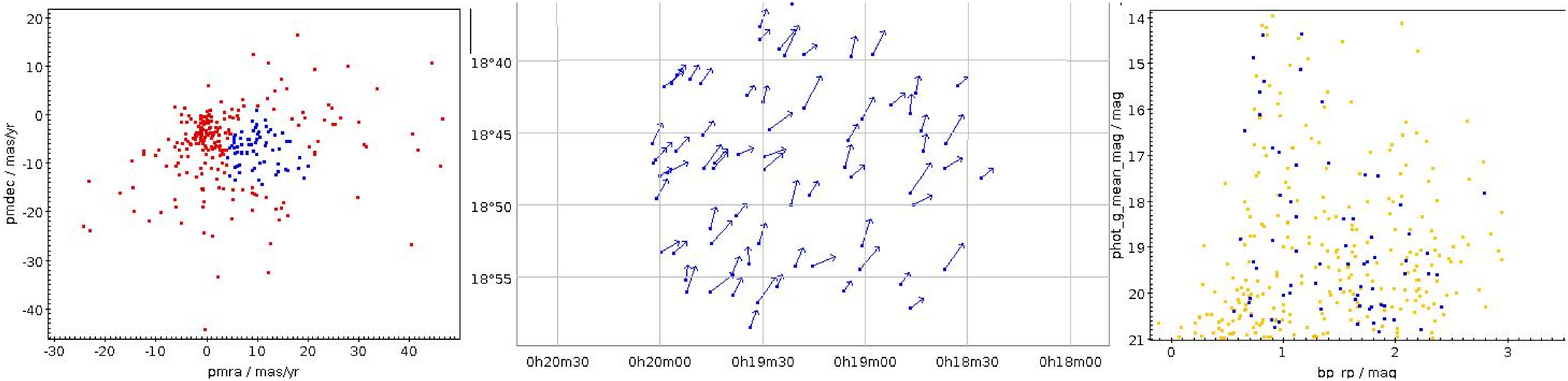}
\caption{Spatial distribution of the ECs in this study (green circles) compared to ECs in  previous works (red circles). Credit: Robert Hurt (NASA/JPL) and Camargo et al.. (2015).}
\label{c438}
\end{figure} 
  
Figure \ref{c438} shows the proper motion plot, probable members and the color magnitude diagram for C438 using Gaia DR2. However, the analysis is not fully conclusive because of the uncertainities, but our impression is that the groups are not real clusters (Private communication with Carme Jordi).
Hence the problem of existence of these clusters still remains unsolved.

\begin{acknowledgements}
The author would like to thank Bhargavi S G for the help in preparation of the proposal submitted to APPS.
\end{acknowledgements}

\end{document}